\documentstyle[prd,preprint,aps]{revtex}
\tighten
\begin{document}
\baselineskip=14pt

\draft

\title{ Gravitational lensing on the Cosmic Microwave Background by 
gravity waves}
\author{Silvia Mollerach}
\address{Departamento de Astronomia y Astrof\'\i sica, 
Universidad de Valencia, E-46100 Burjassot, Valencia, Spain}


\maketitle

\begin{abstract}
We study the effect of a stochastic background of gravitational waves
on the gravitational lensing of the Cosmic Microwave Background (CMB)
radiation. It has been shown that matter density inhomogeneities produce
a smoothing of the acoustic peaks in the angular power spectrum
of the CMB anisotropies. A gravitational wave background gives rise
to an additional smoothing of the spectrum. For the most simple case of a
gravitational wave background arising during a period of inflation, the
effect results to be
three to four orders of magnitude smaller than its scalar counterpart, 
and is thus undetectable. It could play a more relevant role in models 
where a larger background of gravitational waves is produced.
\end{abstract}


The gravitational lensing effect of matter density inhomogeneities 
on the Cosmic Microwave Background (CMB) radiation has been the 
subject of several studies \cite{bl87,co89,to89,li90,ca93,se96,mu96,ma97}. 
It is well known that the deflections undergone by the photons along 
their path since last scattering can modify the pattern of the 
observed anisotropies. The effect is to smooth the acoustic or Doppler 
peaks in the angular spectrum.  Although the effect has been found 
to be small, it should be observable in small angle high 
accuracy observations \cite{se96,ma97}.
It has recently been pointed out \cite{mo97} that a stochastic
background of gravitational waves also contributes to the
gravitational lensing of the CMB radiation.

Many scenarios of the early universe may have produced a 
stochastic background of gravitational waves, as for instance
a period of inflation, phase transitions leading to topological 
defects \cite{vi94}, or bubbles nucleated in a first order phase
transition \cite{ko92}. 
In many inflationary models, the background 
of gravitational waves gives a substantial contribution to the
CMB anisotropies at large angular scales \cite{lu92}. These 
anisotropies arise due to the redshift of the photons induced by 
the time variation of the graviational waves amplitude along the 
photon paths.

In this paper we want to quantify the effect on the CMB 
anisotropies induced by the gravitational lensing of photons from a 
gravitational wave background. We consider a perturbed flat 
Robertson--Walker spacetime described in the Poisson gauge by
\begin{equation}
ds^2= a^2(\eta)\left(-(1+2\varphi)d\eta^2
+\left[(1-2\varphi)\delta_{ij}+\chi^{\top}_{ij}\right]
dx^i dx^j\right),
\end{equation}
where $\eta$ is the conformal time,
in the absence of vector perturbations. $\varphi$ is the peculiar
gravitational potential and $ \chi^{\top}_{ij}$ denotes the tensor
(transverse and traceless) perturbation. 

The gravitational lensing effect on photons is described by the
angular displacement $\vec \beta$ that measures the 
difference between the angular direction on the sky from which a given 
photon arrives to the observer and the one it would have had in the
absence of lensing sources along its path. It is given by
$\vec \beta=r_{\cal{E}}^{-1}{\bf x}_\bot (\lambda_{\cal{E}})$,
with $r_{\cal{E}}=(\eta_{\cal{O}}-\eta_{\cal{E}})$ the distance to
the last scattering surface and
\begin{eqnarray}
x^i_\bot (\lambda_{\cal{E}})
&=&(\delta^{ij}-e^i e^j)\int_{\lambda_{\cal{O}}}^{\lambda_{\cal{E}}} 
d\lambda \left(\chi^{\top}_{jk}e^k-\chi^{\top}_{{\cal{O}}jk}e^k\right)
\nonumber\\
&-&(\delta^{ij}-e^i e^j)\int_{\lambda_{\cal{O}}}^{\lambda_{\cal{E}}} 
d\lambda
(\lambda_{\cal{E}}-\lambda)\left(2 \varphi_{,j}
-\frac{1}{2}\chi^{\top}_{kl,j}e^k e^l
\right),\label{glens}
\end{eqnarray}
where $\hat{\bf e}$ is a unit vector denoting the direction of
arrival of the photons, and the integration is along the photon 
background geodesics parametrized by $\lambda$. The subscript
${\cal{O}}$ denotes quantities evaluated at the observation point
and ${\cal{E}}$
at the emission (or last scattering surface). The term including 
$\varphi$ corresponds to the displacement due to scalar density 
perturbations that has been considered in some previous studies 
\cite{se96,mu96,ma97} and has an observable effect on small angular scales,
while the rest describes the effect of the gravitational wave 
background.

It has been shown by the studies of scalar gravitational lensing that
the effect on the CMB anisotropies can be obtained from the 
autocorrelation function of the transverse displacement
\begin{eqnarray}
S(\alpha)&=&\langle \beta^j(\hat{\bf e}_1) 
\beta_j(\hat{\bf e}_2)\rangle_{(\hat{\bf e}_1 \cdot
\hat{\bf e}_2=\cos\alpha)}
\nonumber\\
&=&\int \frac{d\Omega_{\hat{\bf e}_1}}{4\pi}
\int \frac{d\Omega_{\hat{\bf e}_2}}{2\pi}
\delta(\hat{\bf e}_1 \cdot \hat{\bf e}_2-\cos\alpha)
\langle \beta^j(\hat{\bf e}_1) 
\beta_j(\hat{\bf e}_2)\rangle,
\label{sa}
\end{eqnarray}
where we have taken the mean over all directions separated by an angle 
$\alpha$.
Once this correlation function is known, we can compute the effect on 
the temperature correlation function using the methods developed in ref.
\cite{ca93,se96,ma97} for small angular scales, or that in ref. \cite{mu96}
that apply to arbitrary angular scales.

The gravitational wave background in a flat universe can be 
decomposed as
\begin{equation}
\chi_{ij}^{\top}({\bf x},\eta)=\frac{1}{(2\pi)^3} \int d^3{\bf k}
\exp(i{\bf k}\cdot{\bf x}) \chi_\sigma({\bf k},\eta) 
\epsilon^{\sigma}_{ij}(\hat{\bf k}),
\end{equation}
where $\epsilon^{\sigma}_{ij}(\hat{\bf k})$ is the polarization tensor,
with $\sigma$ ranging over the polarization components $+,\times$, and 
$\chi_\sigma({\bf k},\eta)$ are the corresponding amplitudes. 
The time evolution
of the amplitude during the matter dominated era can be written as
\begin{equation}
\chi_\sigma({\bf k},\eta) \approx A(k) a_\sigma({\bf k}) 
\left(\frac{3 j_1(k\eta)}{k\eta}\right),
\end{equation}
where $a_\sigma({\bf k})$ is a zero mean random variable with 
autocorrelation function 
$\langle a_\sigma({\bf k}) a_{\sigma'}({\bf k'})\rangle =(2\pi)^3 
k^{-3} \delta^3({\bf k}+{\bf k'}) \delta_{\sigma\sigma'}$, 
and $j_1(x)$ denotes the spherical Bessel function of
first order.
The spectrum of the gravitational wave background depends on the 
processes by which it was generated.

For a wave propagating in the direction $\hat{\bf k}$, defining a 
right-handed triad given by $(\hat{\bf k},\hat{\bf m},\hat{\bf n})$,
the polarization tensor can be written as
\begin{eqnarray}
\epsilon^{+}_{ij}(\hat{\bf k})&=& m_i m_j-n_i n_j
\nonumber\\
\epsilon^{\times}_{ij}(\hat{\bf k})&=& m_i n_j +n_i m_j.
\end{eqnarray}

In order to compute the autocorrelation function of the angular
displacement $S(\alpha)$ induced by the gravitational wave background, 
we split $x_{\bot}^i=x_{\bot}^{{\rm I} i}+x_{\bot}^{{\rm II} i}$ with 
\begin{eqnarray}
x_{\bot}^{{\rm I} i}=-\frac{1}{2}(\delta^{ij}-e^i e^j)
\int_{\lambda_{\cal{O}}}^{\lambda_{\cal{E}}} d\lambda
(\lambda_{\cal{E}}-\lambda) \chi^{\top}_{kl,j}e^k e^l,
\nonumber\\
x_{\bot}^{{\rm II} i}=(\delta^{ij}-e^i e^j) 
\int_{\lambda_{\cal{O}}}^{\lambda_{\cal{E}}} d\lambda
\left(\chi^{\top}_{jk}e^k-\chi^{\top}_{{\cal{O}}jk}e^k\right).
\end{eqnarray}
We thus obtain $S(\alpha)=S^{\rm I}(\alpha)+S^{\rm II}(\alpha)$,
with $S^{{\rm I}({\rm II})}(\alpha)=r_{\cal{E}}^{-1}\langle 
x^{{\rm I}({\rm II})j}_\bot 
x^{{\rm I}({\rm II})}_{\bot j}\rangle$, as $\langle x^{{\rm I} j}_\bot 
x^{{\rm II}}_{\bot j}\rangle=0$.
The expressions for $S^{\rm I}(\alpha)$ and $S^{\rm II}(\alpha)$ 
can be obtained
replacing in eq. (\ref{sa}), the above expressions for 
$x_{\bot}^{{\rm I} i}$ and $x_{\bot}^{{\rm II} i}$. 
Parametrizing the photon 
geodesics by $\lambda\equiv(\eta_{\cal{O}}-\eta)/(\eta_{\cal{O}}
-\eta_{\cal{E}})$, so that $\lambda_{\cal{O}}=0$ and
$\lambda_{\cal{E}}=1$, and ${\bf x}=\hat{\bf e} \lambda
(\eta_{\cal{O}}-\eta_{\cal{E}})$, we obtain
\begin{eqnarray}
S^{\rm I}(\alpha)&=&\frac{9}{2}
\int_0^\infty \frac{d\omega}{\omega}\frac{A^2(\omega)}{(2\pi)^3}
\int_0^1 d\lambda \int_0^1 d\lambda' j_1(\omega(1-\lambda))
j_1(\omega(1-\lambda')) 
\nonumber\\
&&\times\int_{-1}^1 d\cos\theta \int_0^\pi d\phi \exp\left({\rm i} \omega
\left[\lambda \cos\theta-\lambda'(\cos\theta \cos\alpha-
\cos\phi \sin\theta \sin\alpha)\right]\right) 
\nonumber\\
&&\times(1-\cos^2\theta+
\cos\phi \sin\theta \sin\alpha \cos\theta \cos\alpha-
\cos^2\phi \sin^2\theta \sin^2\alpha)
\nonumber\\
&&\times\left(2 \cos^2\alpha-1-3\cos^2\theta \cos^2\alpha+
\cos^4\theta \cos^2\alpha+2\cos\phi \sin^3\theta \sin\alpha 
\cos\theta \cos\alpha\right.\nonumber\\
&&\left.+\cos^2\theta+\cos^2\phi \sin^2\alpha
(1-\cos^4\theta)\right),
\label{si}
\end{eqnarray}
where we have defined $\omega\equiv k(\eta_{\cal{O}}-\eta_{\cal{E}})$,
and
\begin{eqnarray}
S^{ \rm II}(\alpha)&=&18
\int_0^\infty \frac{d\omega}{\omega}\frac{A^2(\omega)}{(2\pi)^3}
\int_0^1 d\lambda \int_0^1 d\lambda'\int_{-1}^1 d\cos\theta
\left( \exp({\rm i} \omega \lambda \cos\theta)
\frac{j_1(\omega(1-\lambda))}{\omega(1-\lambda)}
-\frac{j_1(\omega)}{\omega}\right)
\nonumber\\
&&\times\int_0^\pi d\phi \left(\exp[-{\rm i} \omega \lambda'(\cos\theta
\cos\alpha-\cos\phi \sin\theta \sin\alpha)]
\frac{j_1(\omega(1-\lambda'))}{\omega(1-\lambda')}
-\frac{j_1(\omega)}{\omega}\right)
\nonumber\\
&&\times\left(\left[\cos^2\theta-(\cos\theta
\cos\alpha-\cos\phi \sin\theta \sin\alpha)^2\right]\left[
\cos\alpha \sin^2\theta +\cos\phi \sin\theta \sin\alpha \cos\theta
\right]\right.
\nonumber\\
&&+\cos\alpha\left[2 \cos^2\alpha-1-3\cos^2\theta \cos^2\alpha+
\cos^4\theta \cos^2\alpha\right.\nonumber\\
&&+\left.\left. 2\cos\phi \sin^3\theta \sin\alpha 
\cos\theta \cos\alpha+\cos^2\theta+\cos^2\phi \sin^2\alpha
(1-\cos^4\theta)\right]\right),
\label{sii}
\end{eqnarray}

The integrals over $\phi$ in the previous equations can be performed 
analytically in terms of Bessel functions, we do not show the result 
here. We will instead directly show the result of the numerical 
integration over the remaining variables for the case of a background 
of gravitational waves generated during an inflationary period.
In this case, the spectrum is nearly scale invariant and proportional 
to the Hubble constant during inflation. Different 
inflationary models predict slightly different spectral tensor 
index $n_T$ and amplitude of tensor modes. For definiteness, we will 
consider a scale invariant spectrum $n_T= 0$.  An upper limit to the 
amplitude is set by the COBE measurement of CMB anisotropies at
large scales. We can thus write the spectrum as $A^2(k)/(2\pi)^3= 
6\times 10^{-11} T^2(k)$, were we have included the transfer function
for gravitons $T^2(k)$ that takes into account the difference in the 
evolution for modes that entered the horizon during the radiation and
matter dominated eras and can be fitted by  
$T^2(k)=1+1.34(k/k_{eq})+2.5(k/k_{eq})^2$ \cite{tu93}, 
where $k_{eq}$ is the scale 
that entered the horizon at the equality time.

A useful quantity is the dispersion of the difference of the photons 
displacement in two directions defined by
\begin{equation}
\sigma^2(\alpha)=\frac{1}{2}\langle ({\bf \beta}(\hat{\bf e}_1)- 
{\bf \beta}(\hat{\bf e}_2))^2\rangle_{(\hat{\bf e}_1 \cdot
\hat{\bf e}_2=\cos\alpha)}=S(0)-S(\alpha).
\end{equation}
Figure 1 shows the ratio $\sigma (\alpha)/\alpha$ for a range of 
angular separations. As the gravitational
wave background and the peculiar gravitational potential are
two uncorrelated fields, the gravitational lensing displacements 
that they produce are also uncorrelated. Thus, we can describe the 
total effect of gravitational lensing as the sum of the scalar and 
tensor contributions, i. e.
$S_{TOT}(\alpha)=S_S(\alpha)+S_T(\alpha)$ and
$\sigma_{TOT}^2(\alpha)=\sigma_S^2(\alpha)+\sigma_T^2(\alpha)$,
where $S_T(\alpha)$ and $\sigma_T^2(\alpha)$ are the ones
computed in this paper and $S_S(\alpha)$ and $\sigma_S^2(\alpha)$
have been computed by several authors \cite{se96,mu96,ma97}.

It has been shown that the temperature autocorrelation function 
including the effects of gravitational lensing $\tilde C(\theta)$ 
can be obtained from that in the absence of gravitational lensing 
$C(\theta)$ if $S(\theta)$ is known. For small angular scales the 
gravitational lensing effect is to smooth the autocorrelation 
function as if the smoothing was produced by a Gaussian antenna
of width $\sigma(\theta)$.

Comparing the results in Fig. 1 with the corresponding ones
for scalar perturbations in refs. \cite{se96,mu96,ma97}, we see that the 
dispersion of the graviational lensing displacements induced by 
gravitational waves is 3 to 4 orders of magnitude smaller than
the corresponding scalar ones. We thus expect that
$\sigma_{TOT}(\alpha)\simeq\sigma_S(\alpha)$
and that the effect of the gravitational lensing by the gravitational
waves background be undetectable at small scales. This result is 
essentially due to the fact that the amplitude of the gravitational waves 
decreases during the matter dominated era for wavelengths smaller than the
Hubble radius, and thus their contribution at small scales is supressed.
A recent analysis of the effect of a stochastic background of 
gravitational waves on multiple images and weak gravitational lensing has 
found a comparable supression with respect to the corresponding scalar
perturbations lensing \cite{ba96}.

We could wonder if the very large scale modes, that are the larger 
amplitude ones, can lead to an observable effect at large angular scales.
This is not the case as the effects of gravitational lensing are not 
evident at scales for which the angular spectrum is smooth. An estimation 
of this effect using the method proposed in ref. \cite{mu96} with 
the correlation $S(\alpha)$ computed in this paper shows that the effect 
on the temperature anisotropy correlation function at large angular
scales can be of the same order of magnitude than that at small angular 
scales (on the contrary, for scalar perturbations the effect at large 
angular scales 
is much smaller than the small scales one). This is however too 
small to be detectable, besides the fact that cosmic variance at large 
angular scales make small variations in the predicted spectrum untestable.

The results discussed above have been obtained for a gravitational wave 
background produced during a period of inflation. There are however
other scenarios of the early universe in which a larger background of 
gravitational waves is expected. This is the case for example for 
models with  cosmic strings \cite{vi94}
or first order phase transitions \cite{ko92}. 
The method developed in this paper can be 
applied to any of these cases just by replacing the corresponding 
spectra $A(\omega)$ in eqs. (\ref{si}) and (\ref{sii}). It should be taken 
into account that also the temperature anisotropies and the scalar 
gravitational lensing dispersion may be different in alternative theories.

\acknowledgements
It is a pleasure to thank M. Portilla and S. Matarrese for useful
comments and suggestions.
I would like to acknowledge the Vicerrectorado de investigaci\'on
de la Universidad de Valencia for financial support, and the Theory 
Division at CERN for hospitality.

\bigskip
Figure 1: $\sigma(\alpha)/\alpha$ vs. $\alpha$ for a range of $\alpha$
in units of radians.
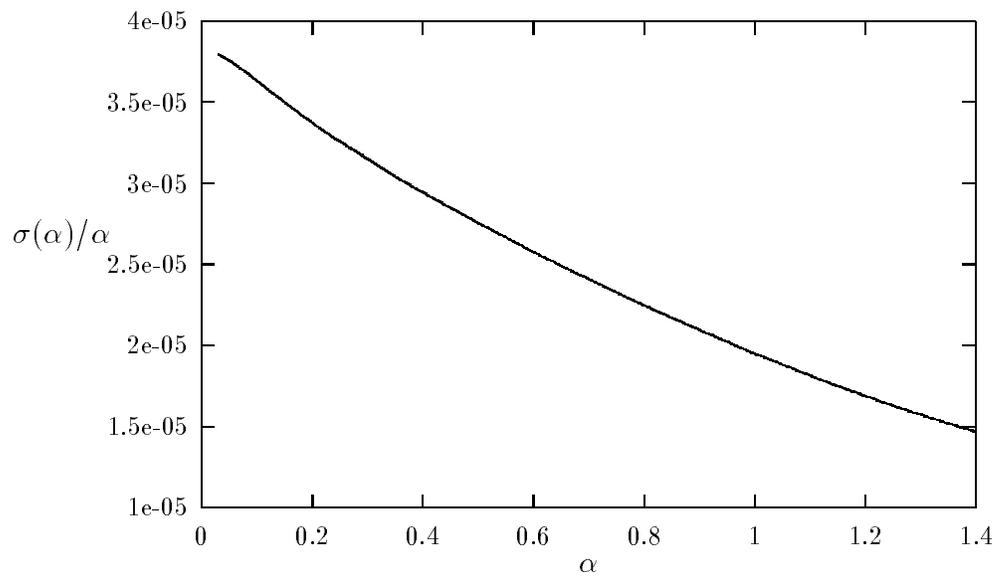
\end{document}